\documentclass[twocolumn,preprintnumbers,amsmath,amssymb]{revtex4}

\usepackage{graphicx}
\usepackage{dcolumn}
\usepackage{bm}

\begin{document}
\title{Comment on ``Radial dependence of radiation
from a bounded source'' by Kirk T.\ McDonald}
\author{A.~Ardavan$^1$, H.~Ardavan$^2$, and J.~Singleton$^3$}
\affiliation{$^1$Clarendon Laboratory, Department of Physics, 
University of Oxford, Parks Road, Oxford~OX1~3PU, United Kingdom\\
$^2$Institute of Astronomy, University of Cambridge,
Madingley Road, Cambridge CB3~0HA, United Kingdom\\
$^3$National High Magnetic Field Laboratory, MS-E536, Los Alamos National Laboratory, 
Los Alamos, New Mexico 87545, USA}
\begin{abstract}
The purpose of this note is to point out that 
McDonald's criticism of our work~\cite{r1} 
is based on a circular argument.  
In order to show that the field of a 
bounded source falls off as $f(\theta,\phi)/r$ 
in the far zone, McDonald uses a 
Huygens-Kirchhoff diffraction 
integral whose derivation (from Maxwell's equations) 
already entails assuming a fall-off of this form 
for the field at infinity~\cite{r2}.  
($r$, $\theta$ and $\phi$ are the spherical polar 
coordinates centred on a point within the source, 
and $f$ is a factor independent of $r$.)
\end{abstract}
\maketitle
Green's theorem can be used to express a scalar field (e.g.\ a component of the electric or magnetic field) inside an empty closed volume $V$ in terms of the values of the field and its normal derivative on the surface $S$ bounding $V$.  The result, when the field $\psi({\bf x},t)$ has a harmonic time dependence $\exp(-i\omega t)$, is
\begin{equation}
\psi({\bf x})=-{1\over 4\pi}\oint_S{e^{ikR}\over R}{\bf n}^\prime\cdot\Big[{\bf\nabla}^\prime\psi+ik\Big(1+{i\over kR}\Big){{\bf R}\over R}\psi\Big]da^\prime,
\label{e1}
\end{equation}
where ${\bf R}={\bf x}-{\bf x}^\prime$, ${\bf x}$ and ${\bf x}^\prime$ are the position vectors of the observation point and boundary points, respectively, $k=\omega/c$, and $c$ is the speed of light {\it in vacuo}; see equation (10.79) of Jackson's {\it Classical Electrodynamics}~\cite{r2}.\par
This is not a solution to Maxwell's equations but just an identity.  As emphasized by Jackson, to apply Eq.~\ref{e1}, ``it is necessary to know the values of $\psi$ and ${\bf n}\cdot{\bf\nabla}\psi$ on the surface $S$.  Unless the problem has been solved by other means, these values are not known'' (page 480 of~\cite{r2}).  Nor can one prescribe the values of both these quantities on $S$: the well-posed boundary conditions for the Helmholtz wave equation (which is an equation of the elliptic type) consist of specifying either $\psi$ (the Dirichlet boundary condition) or ${\bf n}\cdot{\bf\nabla}\psi$ (the Neumann boundary condition) on a closed surface (see page 37 of~\cite{r2}).\par
In the diffraction context, it is assumed that the boundary $S$ consists of two surfaces ($S_1$ and $S_2$) one of which ($S_2$) lies in the far zone.  The integral over $S$ is thus divided into two parts, one over the screen and its apertures ($S_1$), the other over a surface at infinity ($S_2$).  The field and its derivative at infinity are then assumed to satisfy
\begin{equation}
\psi\to f(\theta,\phi){e^{ikr}\over r},\qquad\qquad {1\over\psi}{\partial\psi\over\partial r}\to \Big(ik-{1\over r}\Big),
\label{e2}
\end{equation} 
where $r=\vert{\bf x}\vert$; see equation (10.78) of~\cite{r2}.  {\it These are the conditions that would hold only for a conventional, spherically spreading radiation; there is no reason to assume that they should hold for all emissions.}  Under these conditions, the integral over $S_2$ vanishes (see~\cite{r2}) and the field can be expressed in terms of its value on $S_1$, as in equation (1) of McDonald~\cite{r1}.  Thus, the equation on which McDonald bases his argument does not follow from Maxwell's equations unless one already assumes that the field decays like $1/r$ for $r\to\infty$.\par
To derive the characteristics of the nonspherically decaying radiation beam that is generated in the experiments reported in~\cite{r3}, it is essential that one takes account of retardation effects~\cite{r4,r5}, i.e.\ that one solves Maxwell's equations with a source term and with Cauchy's boundary conditions (see page 38 of~\cite{r2}).  Equation~\ref{e1} is identically satisfied once one inserts the values of $\psi$ and ${\bf n}\cdot{\bf\nabla}\psi$ from the retarded solution of the wave equation into the kernel of the exact Huygens-Kirchhoff diffraction integral. The asymptotic conditions expressed in Eq.~\ref{e2}, on the other hand, do not hold unless the known solution used for obtaining the values of $\psi$ and ${\bf n}\cdot{\bf\nabla}\psi$ on $S$ describes a conventional, spherically spreading radiation.  In the case of the nonspherically spreading radiation generated by a rotating superluminal source, for instance, the asymptotic value of $\psi$ is proportional to $r^{-1/2}$, rather than $r^{-1}$, for large $r$. \par
\begin{figure}[tbp]
   \centering
\includegraphics[width=8cm]{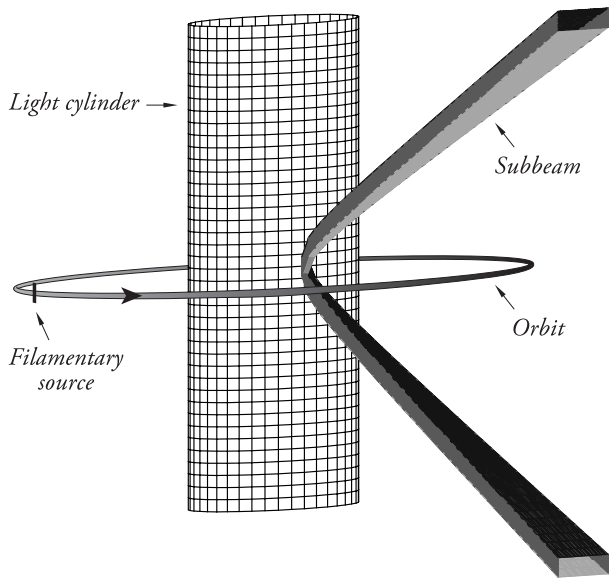}
\caption{Schematic illustration of the light cylinder (the surface on which a rigidly rotating extended source attains the speed of light),
the filamentary part of the source
that approaches the observeration point
with the speed of light and zero acceleration
at the retarded time,
the orbit of this filamentary source,
and the subbeam formed by the caustics
that emanate from the constituent volume elements of this filament. The figure represents a snapshot
corresponding to a fixed value of the observation time. 
The subbeam is diffractionless in the polar direction.
The polar width of this subbeam
decreases with the distance $r$
in such a way that the thickness
of the subbeam in the polar direction
remains constant: 
it equals the projection
of the length
of the contributing filamentary source
onto a direction normal to the line of sight. 
The azimuthal width of the subbeam,
on the other hand,
is subject to diffraction as in any other radiation beam.}
\label{beam}
\end{figure}

That this nonspherical decay of the field does not contravene conservation of energy has now been explicitly demonstrated in~\cite{r6}.  The part of the superluminal source that makes the main contribution toward the value of the nonspherically decaying field has a filamentary structure whose radial and azimuthal widths become narrower (as $r^{-2}$ and $r^{-3}$, respectively), the farther is the observer from the source.  The loci on which the waves emanating from this filament interfere constructively delineate a radiation subbeam that is nondiffracting in the polar direction (see Fig.~\ref{beam}).  The cross-sectional area of each nondiffracting subbeam increases as $r$, instead of $r^2$, so that the requirements of the conservation of energy are met by the nonspherically decaying radiation automatically.  The overall radiation beam within which the field decays nonspherically consists, in general, of the incoherent superposition of such nondiffracting radiation subbeams. (For a detailed discussion of these findings, see~\cite{r6}.)\par
A final remark is in order:  the far-field approximation to the retarded potential, given in equation (9.8) of~\cite{r2}, does not support McDonald's argument as claimed in the first footnote of~\cite{r1}.  In situations where caustics occur, it is essential that one should evaluate the radiation integrals prior to proceeding to the far-field limit.  Because it replaces the spherical wave fronts by planar ones, the far-field approximation obliterates significant geometrical features of the loci of stationary points of the phases of the integrands in these integrals (see Section 4B of~\cite{r5}).\par

\end{document}